\documentstyle[prl,aps,preprint,eqsecnum]{revtex}

\begin{document}
 
\title{G\"odel type solution in teleparallel gravity}
 
\author{Yu.N. Obukhov\footnote{On leave from: Department of
Theoretical Physics, Moscow State University, 117234 Moscow, Russia}}
\address{Institute for Theoretical Physics, University of Cologne,
50923 K\"oln, Germany}
\author{T. Vargas}
\address{Instituto de F\'{\i}sica Te\'orica,
Universidade Estadual Paulista,
Rua Pamplona 145,\\
01405-900\, S\~ao Paulo SP,
Brazil}

\maketitle

\begin{abstract}
The stationary cosmological model {\it without} closed timelike curves
of G\"odel type is obtained for the ideal dust matter source within
the framework of the teleparallel gravity. For a specific choice of the 
teleparallel gravity parameters, this solution reproduces the causality
violating stationary G\"odel solution in general relativity, in accordance 
with the teleparallel equivalent of general relativity. The relation 
between the axial-vector torsion and the cosmic vorticity is clarified. 
\end{abstract}


\section {Introduction}

Cosmological models with rotation attract considerable attention since
the early papers of G\"odel \cite{gd}. However, there are two fundamental
difficulties of the original G\"odel rotating model which made its physical
significance problematic: the absence of expansion and the presence of
the closed timelike curves. The latter is equivalent to the possibility
of violation of causality in such a spacetime. Much of the research on 
this subject \cite{hwe,reb,pfarr,mo,os,bn,nr,l,ac,hn,wi,am,ls,ok,aman} 
was aimed at removing both
drawbacks and thereby in constructing a realistic cosmology with rotation
and expansion. An overview of the corresponding developments and a
comprehensive list of the relevant references is given in \cite{rotrev}.
 
An alternative approach to gravitation is the so called teleparallel gravity 
(see \cite{M,mhn,mg9} where the basic references on the relevant literature 
can be found) which corresponds to a gauge theory for the translation
group based on the Weitzenb\"ock geometry~\cite{we}. In this theory,
gravitation is attributed to torsion~\cite{xa}, which plays the role of a
force~\cite{pe}, whereas the curvature tensor vanishes identically. Such
a spacetime (Weitzenb\"ock manifold) is characterized by a
nontrivial tetrad field as the fundamental structure, and this field gives
rise to the metric as a by-product. The translational gauge potentials
appear as a nontrivial part of the tetrad field, and they induces on
spacetime a teleparallel structure which is directly related to the
presence of the gravitational field.
 
In this work we address the above mentioned problems of a rotating 
world in the framework of the teleparallel gravity theory. The general 
teleparallelism model is characterized by the three coupling constants, 
and as a result the structure of the corresponding field equations is 
less rigid than that of the general relativity theory. Using this fact, 
one can expect that there exist classes of teleparallel models which are 
free of one or both difficulties.

More specifically, in the current paper we demonstrate that the completely
causal cosmological G\"odel type solutions do exist in the appropriate
teleparallel gravity models.
 
The structure of the paper is as follows. In Section~\ref{sec1}, we review
the main features of general teleparallel gravity. In Section~\ref{sec2}, 
we obtain the teleparallel G\"odel type solution. Conclusions are presented 
in Section~\ref{conclusions}.

\section{Teleparallel Gravity}\label{sec1}

In the teleparallel gravity, the spacetime is represented by the
Weitzenb\"{o}ck manifold $W_{4}$ of distant parallelism. This gravitational
theory naturally arises within the gauge approach based on the group of the 
spacetime translations. In simplified terms (see \cite{pbo,percacci,gron,rt} 
for the advanced discussion of gauge theory of gravity in the framework 
of nonlinear connections), this approach introduces at each point of the 
spacetime manifold a tetrad (coframe) field $h^{a}{}_{\mu}$. Its inverse 
(frame) $h_{a}{}^{\nu}$ satisfies the condition $h^{a}{}_{\mu}\,h_{a}{}^{\nu}
=\delta^{\nu}_{\mu}$. The tetrad field induces a teleparallel structure on 
the spacetime which is directly related to the gravitational field, and the
Riemannian metric arises as
\begin{equation}
g_{\mu\nu} = \eta_{ab}\;h^a{}_\mu\;h^b{}_\nu\;,\label{met}
\end{equation}
with the Minkowski matrix $\eta_{ab} = {\rm diag}(+1,-1,-1,-1)$. A parallel 
transport of a tetrad $h^{a}{}_{\mu}$ between two neighbouring points is 
encoded in the covariant derivative
\begin{equation}
\nabla_{\nu}h^{a}{}_{\mu} = \partial_{\nu}h^{a}{}_{\mu}
-\Gamma ^{\alpha}{}_{\mu\nu}h^{a}{}_{\alpha},\label{nt}
\end{equation}
where $\Gamma^{\alpha}{}_{\mu \nu}$ is the Weitzenb\"ock connection.
Imposing the geometric condition that the tetrads are parallelly
transported in the Weitzenb\"ock space-time, we obtain
\[
\nabla_{\nu}h^{a}{}_{\mu}=\partial_{\nu}h^{a}{}_{\mu}
- \Gamma^{\alpha}{}_{\mu \nu}h^{a}{}_{\alpha}\equiv 0.
\]
This is the condition of the absolute parallelism, or teleparallelism.
It is equivalent to
\begin{equation}
\Gamma^{\alpha}{}_{\mu\nu}=h_{a}{}^{\alpha}\partial_{\nu}h^{a}{}_{\mu}
\end{equation}
which gives the explicit form of the Weitzen\-b\"ock connection in
terms of the tetrad. The antisymmetric part of connection,
\begin{equation}
T^{\rho}{}_{\mu\nu} = \Gamma^{\rho}{}_{\nu\mu} - \Gamma^{\rho}{}_{\mu\nu},
\end{equation}
is the torsion tensor of the Weitzenb\"ock connection. The curvature of 
the Weitzenb\"ock connection vanishes identically as a consequence of 
absolute parallelism.
 
It has been shown in~\cite{xa} that the most general action in the
teleparallel gravity coupled to the matter has the form
\begin{equation}\label{action}
S = \int d^4x\,h\left[\frac{1}{4\kappa}\,(L + \Lambda) + L^{\rm mat}\right],
\end{equation}
with $h = \det\,h^a{}_\mu$, and
\begin{equation}
L = c_1\,T^{\lambda\tau\nu}T_{\lambda\tau\nu} + c_2\,T^{\lambda\tau\nu}
T_{\nu\tau\lambda} + c_3\,T^{\sigma\lambda}{}_{\sigma}
T^{\sigma}{}_{\lambda\sigma}.\label{lag}
\end{equation}
Here $c_1$, $c_2$, and $c_3$ are the three dimensionless coupling
constants, $\Lambda$ is a cosmological term, and $\kappa$ is the Einstein
gravitational constant. For the specific choice of the parameters,
\begin{equation}
c_1=\frac{1}{4},\qquad c_2=\frac{1}{2},\qquad c_3 = -\,1,
\label{par}
\end{equation}
the teleparallel gravity reduces to the so called teleparallel equivalent
of the general relativity. The Lagrangian (\ref{lag}) may be rewritten as
\begin{equation}
L = S^{\lambda\tau\nu}\,T_{\lambda\tau\nu},
\end{equation}
where
\begin{equation}
S^{\lambda\tau\nu} = c_1T^{\lambda\tau\nu} + \frac{c_2}{2}\left(
T^{\tau\lambda\nu} - T^{\nu\lambda\tau}\right) + \frac{c_3}{2}\left(
g^{\lambda \nu} \;T^{\sigma\tau}{}_{\sigma} - g^{\tau \lambda} 
\;T^{\sigma\nu}{}_{\sigma}\right).\label{2.32a}
\end{equation}
In the present paper, we use the traditional tensor formalism. Those
readers, who prefer a more advanced exterior calculus approach, should
address \cite{opp} for a general discussion of teleparallel gravity in
the metric-affine approach. Here, we only mention that the gravity
field Lagrangian can be rewritten in terms of the quadratic contractions
of the three irreducible torsion pieces with the corresponding coupling 
constants $a_1, a_2, a_3$. Then one can verify that the original coupling 
constants $c_1, c_2, c_3$ are related to the new parameters as follows: 
$c_1 = (2a_1 + a_3)/3$, $c_2 = 2(a_2 - a_1)/3$, and $c_3 = 2(a_3 - a_1)/3$, 
see \cite{opp}. 

By performing variation of the action (\ref{action}) with respect to 
$h^{a}{}_{\mu}$, we get the teleparallel gravitational field equation
\begin{equation}
{\frac 1 h}\,\partial_\sigma\left(h S_{\lambda}{}^{\rho\sigma}\right) 
+ S_\nu{}^{\mu\rho}\,\Gamma^\nu{}_{\mu\lambda}- \frac{1}{4}
\,\delta^\rho_\lambda\,L = \kappa\,T_{\lambda}{}^{\rho} 
+ \Lambda\,\delta^{\rho}_{\lambda},\label{eqc}
\end{equation}
where $T_{\lambda}{}^{\rho} = h^a{}_\lambda\,\delta (hL^{\rm mat})/(h\delta
h^a{}_\rho)$ is the energy-momentum tensor of matter. In the rest of the
paper we confine our attention to the case of the ideal fluid with
\begin{equation}\label{fluid}
T_{\lambda}{}^{\rho} = (\varepsilon + p)\,u_{\lambda}u^{\rho} 
- p\,\delta^{\rho}_{\lambda}.
\end{equation}
Here $\varepsilon$ is the energy density, $p$ is pressure, and $u^\mu$ is
the fluid 4-velocity (normalized by $u^\mu u_\mu = 1$).

\section{The teleparallel G\"odel type solution}\label{sec2}

The line element of the G\"odel type cosmological model is given by the
interval
\begin{equation}
ds^{2}= dt^{2} - 2\sqrt{\sigma}\,a(t)e^{mx}\,dtdy 
- a(t)^{2}(dx^{2} + k\,e^{2mx}dy^{2} + dz^{2}),\label{dsi}
\end{equation}
where $m$, $\sigma$, $k$ are the constant parameters and $a(t)$ is the
time-dependent cosmological scale factor. A general analysis \cite{mo,rotrev}
of the kinematic properties of such a spacetime shows that the closed 
timelike curves are absent in this manifold when $k>0$ ($\sigma> 0$ by
definition). 
 
Using the relation (\ref{met}) we obtain the tetrad:
\begin{equation}
\label{te1}
\widetilde{h}^{a}{}_{\mu} =  \pmatrix{
1 & 0 & -\sqrt{\sigma}\,a\,e^{mx}  & 0 \cr
0 & a & 0  & 0 \cr
0 & 0  & a\sqrt{\sigma + k}\,e^{mx} & 0 \cr
0 & 0  & 0 & a},
\end{equation}
and its inverse is
\begin{equation}
\label{te2}
\widetilde{h}_{a}{}^{\mu} =  \pmatrix{
1 & 0 & 0  & 0 \cr
0 & a^{-1} & 0  & 0 \cr
\sqrt{\frac{\sigma}{k + \sigma}} & 
0  & \frac{e^{-mx}}{a\sqrt{\sigma + k}} & 0 \cr
0 & 0  & 0 & a^{-1}}.
\end{equation}

Using the Lorentz transformation for the tetrad $\widetilde{h}^{a}{}_{\mu}$
\begin{equation}
h^{a}{}_{\mu}= \Lambda{}^{a}{}_{b}\,\widetilde{h}^{b}{}_{\mu},\label{cofr}
\end{equation}
we obtain the Weitzenb\"ock connection
\begin{equation}
\Gamma^{\alpha}{}_{\mu \nu}=\widetilde{\Gamma}^{\alpha}{}_{\mu \nu}+
\widetilde{h}_{b}{}^{\alpha}\;({\Lambda}^{-1}\partial_{\nu}\Lambda)^{b}{}_{c}
\;\widetilde{h}^{c}{}_{\mu}.
\end{equation}
We assume that the Lorentz matrix depends on the cosmological time only.
Hence (denoting the time derivative by the dot) $\partial_{\nu}\,\Lambda^a{}_c
= {\delta}^{t}_{\nu}\,\dot{\Lambda}{}^a{}_c$ and consequently
\begin{equation}
({\Lambda}^{-1}\partial_{\nu}\Lambda)^{b}{}_{c}= {\delta}^{t}_{\nu}
\,L^{b}{}_{c},\label{l1}
\end{equation}
with an antisymmetric matrix ($L_{bc} = -\,L_{cb}$) of the form
\begin{equation}
\label{matr}
L^{b}{}_{c} =  \pmatrix{
0 & v_1 & v_2  & v_3 \cr
v_1 & 0 & -s_3  & s_2 \cr
v_2 & s_3  & 0 & -s_1 \cr
v_3 & -s_2  & s_1 & 0}.
\end{equation}
 
{}From Eqs. (\ref{te1}), (\ref{te2}) and (\ref{matr}), we can now construct
the Weitzenb\"ock connection. Its nonvanishing components read:
\begin{eqnarray}
\Gamma^{t}{}_{tt} = v_2\,\sqrt{\frac{\sigma}{\sigma + k}}\;,\quad
\Gamma^{t}{}_{xt} = a\left(v_1 + s_3\,\sqrt{\frac{\sigma}{\sigma + k}}\right)
\;,\nonumber\\
\Gamma^{t}{}_{yt} = v_2\,{\frac {ae^{mx}k}{\sqrt{\sigma + k}}}\;,\quad
\Gamma^{t}{}_{zt} = a\left(v_3 - s_1\,\sqrt{\frac{\sigma}{\sigma + k}}\right)
\;,\nonumber\\
\Gamma^{x}{}_{tt} = {\frac{v_1}{a}}\;,\quad
\Gamma^{x}{}_{xt} = \Gamma^{z}{}_{zt}= {\frac{\dot{a}}{a}}\;,\quad
\Gamma^{x}{}_{yt}= -\,e^{mx}(s_3\,\sqrt{\sigma + k} + v_1\,\sqrt{\sigma})
\;,\label{ga}\\
\Gamma^{x}{}_{zt} =-\Gamma^{z}{}_{x t}= s_2\;,\quad
\Gamma^{y}{}_{tt} = v_2\,{\frac{e^{-mx}}{a\sqrt{\sigma + k}}}\;,\nonumber\\
\Gamma^{y}{}_{xt} = s_3\,{\frac{e^{-mx}}{\sqrt{\sigma + k}}}\;,\quad
\Gamma^{y}{}_{yt} = {\frac{\dot{a}}{a}} - v_2\,\sqrt{\frac{\sigma}{\sigma + k}}
\;,\quad
\Gamma^{y}{}_{zt} = -s_1\,{\frac{e^{-mx}}{\sqrt{\sigma + k}}}\;,\nonumber\\
\Gamma^{y}{}_{yx} = m \;,\quad \Gamma^{z}{}_{tt} = {\frac{v_3}{a}}\;,\quad
\Gamma^{z}{}_{yt} = e^{mx}(s_1\,\sqrt{\sigma + k} - v_3\,\sqrt{\sigma})
\;.\nonumber
\end{eqnarray}
Here, as before, the dot denotes a derivative with respect to the time $t$.

Assuming that the matter is co-moving with the space, i.e., taking the
velocity $u^\alpha = \delta^\alpha_t$, we find the components of the
energy-momentum tensor (\ref{fluid}):
\begin{equation}
T_t{}^t = \varepsilon,\quad T_x{}^x = T_y{}^y = T_z{}^z = - p,\quad
T_y{}^t = -\,\sqrt{\sigma}\,a\,e^{mx}\,(\varepsilon + p).
\end{equation}

\subsection{Solving the field equations}

In this paper we will study the case of the stationary world with the
constant scale factor $a = const$. In this way we will be able to perform
a direct comparison of the solutions obtained with the G\"odel solution
in the general relativity theory.
 
The value of the parameter $2c_1 - c_2$ is essential. When this is equal
zero, i.e., $2c_1 = c_2$ [note that this case includes the teleparallel
equivalent of general relativity (\ref{par})], then the difference of
the $(_y{}^y)$ and $(_z{}^z)$ components of the field equation (\ref{eqc})
yields:
\begin{equation}
{\frac {(2k + \sigma)\,m\,c_2} {2a^2(k + \sigma)}} = 0.
\end{equation}
In other words, we have three possibilities: either $m = 0$, then the
cosmic vorticity vanishes, or the coupling constant $c_2 = 0$, but then
also $c_1 =0$ and the resulting Lagrangian is unphysical. Finally, when
both $m$ and $c_2$ are nontrivial, we should have $k = -\,\sigma/2$. This
is exactly the value of the parameter $k$ of the original G\"odel model
which contains the closed timelike curves. Accordingly, we will not
consider this case as it basically leads to the qualitatively same model.
 
Thus we instead assume that $2c_1 - c_2 \neq 0$. Then the $(_z{}^t)$-th
component of the field equation yields $s_2 = 0$, whereas $(_x{}^z)$-th
component yields
\begin{equation}
s_1 = v_3\,\sqrt{\frac {\sigma} {\sigma + k}}.\label{s1}
\end{equation}
Furthermore, subtracting $(_y{}^y)$ from $(_z{}^z)$, we find
\begin{equation}\label{s3}
s_3 = - v_1\,\sqrt{\frac {\sigma} {\sigma + k}} + {\frac ma}\,{\frac
{2kc_1 + (k + \sigma)c_2}{(2c_1 - c_2)\sqrt{\sigma}\sqrt{\sigma + k}}}.
\end{equation}
Now, the component $(_t{}^t)$ minus $(_x{}^x)$ plus the product of
$(e^{-mx}/a\sqrt{\sigma})$ times $(_y{}^t)$ yields the equation
\begin{equation}
{\frac {k\sigma(2c_{1} + c_{2} + c_{3})[k(v_{1}^2 + v_{3}^2)
+ (k + \sigma)v_{2}^2] + v_1{\frac ma}Q_1 + 2{\frac {m^2}{a^2}}
\,P_1}{2\sigma (k + \sigma)^2}} = 0.\label{QP1}
\end{equation}
Here we have denoted the following combinations of the constants:
\begin{eqnarray}
Q_1 &=& {\frac {k\sigma (2c_1+c_2+c_3)[2(k -\sigma)c_{1}
+ 3(k + \sigma)c_{2}]} {2c_{1} - c_{2}}},\label{Q1}\\
P_1 &=& [\sigma c_1 - (k + \sigma)c_2]\big[ 4(2k^2 + 2k\sigma + \sigma^2)
\,c_1^2 - 2\sigma(2k + \sigma)\,c_1c_2 \nonumber\\
&& -\,2k\sigma\,c_1c_3 - 2(k + \sigma)^2\,c_2^2 - \sigma(k + \sigma)
\,c_2c_3\big]/(2c_{1} - c_{2})^2.\label{P1}
\end{eqnarray}
The $(_t{}^x)$-th and the $(_t{}^y)$-th components of the field equation
read, respectively:
\begin{eqnarray}
v_{2}\,m\,{\frac {(2c_1 + c_2)[2(k + \sigma)c_1 - (3k + 2\sigma)c_2]
- [2kc_1 + (k + \sigma)c_2]c_3} {2\sqrt{\sigma}\sqrt{k + \sigma}
(2c_{1} - c_{2})a^2}} &=& 0,\label{v2}\\
{\frac {-\,\sigma (2c_{1} + c_{2} + c_{3})[k(v_{1}^2 + v_{3}^2)
+ (k + \sigma)v_{2}^2] - v_1{\frac ma}Q_2 + 2{\frac {m^2}{a^2}}
\,P_2}{2\sqrt{\sigma}\,e^{mx}a(k + \sigma)^2}} &=& 0.\label{QP2}
\end{eqnarray}
Here we have introduced further abbreviations:
\begin{eqnarray}
Q_2 &=& \bigg\{(2c_{1} + c_{2})\left[2k(k + 3\sigma)c_{1} - (3k-\sigma)
(k + \sigma)c_{2}- (k + \sigma)^2c_{3} \right]\nonumber\\ \label{Q2}
&& + [4kc_{1} + 3(k + \sigma)c_{2}]c_{3}\sigma\bigg\}/(2c_{1} - c_{2}),\\
P_2 &=& {\frac {(2c_{1} + c_{2} + c_{3})[2\sigma c_1 + (k + \sigma)c_2]
[(k + \sigma)c_2 - \sigma c_1]}{(2c_{1} - c_{2})^2}}.\label{P2}
\end{eqnarray}
 
Let us analyse the two equations (\ref{QP1}) and (\ref{QP2}) which can be
considered as the algebraic system for $v_1$ and $v_3$. Then discarding the
denominators, from the difference and the sum of the numerators in
(\ref{QP1}) and (\ref{QP2}) we find
\begin{eqnarray}
v_1 = -\,{\frac {2m}{a}}\,{\frac {P_1 + kP_2}{Q_1 - kQ_2}} &=&
{\frac {2m\,[\sigma c_1 - (k + \sigma)c_2]}{a\,k(2c_1 - c_2)}},\label{v1}\\
\sigma (2c_{1} + c_{2} + c_{3})[k(v_{1}^2 + v_{3}^2)
+ (k + \sigma)v_{2}^2] &=& {\frac {2m^2}{a^2}}\,{\frac {P_1Q_2 + P_2Q_1}
{Q_1 - kQ_2}} \nonumber\\ \label{vvv}
&=& -\,{\frac {2m^2P_0\,[\sigma c_1 - (k + \sigma)c_2]}{a^2\,k(2c_1 - c_2)^2}}.
\end{eqnarray}
Here we denoted
\begin{eqnarray}
P_0 &=& 4k(2k + 3\sigma)\,c_1^2 + 2\sigma(2k + \sigma)\,c_1c_2 - \sigma^2
\,c_1c_3 \nonumber\\
&& -\,(k + \sigma)(2k - \sigma)\,c_2^2 + 2\sigma(k + \sigma)\,c_2c_3.
\end{eqnarray}
Note that although the expressions (\ref{Q1}), (\ref{P1}), and (\ref{Q2}),
(\ref{P2}) are quite nontrivial, the final result (\ref{v1})-(\ref{vvv})
is rather simple. Substituting (\ref{v1}) into (\ref{vvv}), we obtain
explicitly
\begin{equation}
v_3^2 = {\frac {k + \sigma}k}\left[ -\,v_2^2 + {\frac {2m^2(2k + \sigma)
\,[(k + \sigma)c_2 - \sigma c_1](2c_1 + c_2)}{a^2\,k\sigma (2c_{1} + c_{2}
+ c_{3})(2c_1 - c_2)}}\right].\label{v3}
\end{equation}
Obviously, we can put $v_2 = 0$ from the equation (\ref{v2}). Otherwise
(\ref{v2}) determines the parameter $k$ in terms of $\sigma$ and $c_1,
c_2, c_3$.

Finally, taking into account (\ref{s1}), (\ref{s3}), (\ref{v1}), and
(\ref{v3}), we can write down the diagonal components of the field
equation (\ref{eqc}):
\begin{eqnarray}\label{ener}
-\,{\frac {3m^2k(2c_1 - c_2)}{2\sigma a^2}} &=& \kappa\,\varepsilon +\Lambda,\\
-\,{\frac {m^2k(2c_1 - c_2)}{2\sigma a^2}} &=& -\,\kappa\,p + \Lambda.\label{p}
\end{eqnarray}
In order to compare with the old G\"odel model, we now specialise to the
case of the dust cosmological matter, $p=0$. Then equation (\ref{p})
determines the parameter $k/\sigma$ in terms of the coupling and geometrical
constants, whereas (\ref{ener}) yields the energy density of the stationarily
rotating world:
\begin{equation}
\varepsilon = {\frac {m^2k(c_2 - 2c_1)}{\kappa\sigma a^2}}.
\end{equation}
Evidently, for the positivity of energy, the teleparallel coupling constants
should satisfy the inequality $c_2 > 2c_1$. The important point is that the
ratio $k/\sigma$ must also be then positive. Hence the resulting rotating
universe does not contain {\it closed} timelike curves, and consequently
the causality is not violated.

\subsection{Finding the Lorentz matrix factor}

In order to complete the description of the exact solution obtained, it
remains to find the Lorentz matrix $\Lambda^a{}_b$ which enters the
tetrad (\ref{cofr}). It should be noted that from the point of view of
the resulting Riemannian geometry, this Lorentz factor is irrelevant
since the spacetime interval (\ref{dsi}) is invariant under the local
Lorentz transformations of the tetrad. However, since we deal with the
Weitzenb\"ock geometry, the Lorentz factor is the essential element of
the definition of the tetrad (\ref{cofr}) and, accordingly, of the
connection (\ref{ga}). Assuming that the elements of the unknown matrix 
$\Lambda^a{}_b$ depend only on cosmological time $t$, we can rewrite 
the differential equation (\ref{l1}) as
\begin{equation}
{\dot\Lambda}^a{}_b = \Lambda^a{}_c\,L^c{}_b\,.\label{dotL}
\end{equation}
Fixing the index $a$, we thus find that the four elements of each row
$\Lambda^0{}_a$, or $\Lambda^1{}_a$, or $\Lambda^2{}_a$, or $\Lambda^3{}_a$
satisfy the system of the linear differential equations (\ref{dotL}) with 
the constant coefficients
\begin{equation}
\label{matr2}
L^a{}_b =  \pmatrix{
0 & v_1 & 0  & v_3\cr
v_1 & 0 & -s_3  & 0 \cr
0 & s_3  & 0 & -s_1\cr
v_3 & 0  & s_1 & 0}.
\end{equation}
The elements of this matrix were determined above in the equations (\ref{s1}),
(\ref{s3}), (\ref{v1}) and (\ref{v3}) and they depend on the coupling
constants of the model and on the geometric parameters of the ansatz.
However, the explicit form of these relations is not needed for our
purposes. The solution of such a system is straightforward: We assume the
standard ansatz of the $\Lambda^b{}_a = e^{\lambda t}\,\alpha^b{}_a$. Then
the constant coefficients $\alpha^b{}_a$ are nontrivial, provided the
characteristic equation is satisfied:
\begin{equation}
\label{cha}
\left|\begin{array}{rrrr}
-\lambda & v_1 & 0 &  v_3 \\
v_1 & -\lambda & s_3 &  0 \\
0 & -s_3 &  -\lambda & s_1 \\
v_3 & 0  & -s_1 & -\lambda
\end{array}\right|=0\,.
\end{equation}
Evaluating the determinant, we find the algebraic equation
\begin{equation}
\lambda^4 + \lambda^2\left(s_1^2+s_3^2-v_1^2-v_3^2\right)
- \left(s_1v_1+s_3v_3\right)^2 = 0.\label{lam4}
\end{equation}
It is easy to see that of the four roots $\lambda_{(A)}$, $A=1,\dots,4$,
two are real and two are purely imaginary. When $\lambda$ is a solution of
the equation (\ref{lam4}), the components $\alpha^a{}_0$ and $\alpha^a{}_1$
represent the independent free parameters, whereas the rest of the
components are expressed as
\begin{equation}
\alpha^a{}_2 = \frac{\lambda^2 - v_1^2 - v_3^2}{v_1s_3 - v_3s_1}\,\alpha^a
{}_0,\qquad \alpha^a{}_3 = \frac{\lambda^2 + s_3^2 - v_1^2}{s_1s_3 + v_1v_3}
\,\alpha^a{}_1\,.\label{alphas}
\end{equation}
The general solution of the system (\ref{dotL}) has the form
\begin{equation}\label{lambda}
\Lambda^a{}_b=\sum_{i=A}^{4}C_{(A)}\alpha^a{}_b\,e^{\lambda_{(A)}t}\,,
\end{equation}
where the $C_{(A)}$ are arbitrary constants. It is thus interesting to
note that whereas the geometry of the solution is stationary, with the
Riemannian curvature and the Weitzenb\"ock torsion tensors being independent
of the cosmological time, the teleparallel tetrad has a nontrivial
time dependence via the Lorentz matrix (\ref{lambda}), in general.

\section{Discussion and conclusion}\label{conclusions}

In this paper we have obtained the teleparallel version of the stationary
rotating cosmological solution of G\"odel type within the framework of the
3-parameter model with the general teleparallel Lagrangian (\ref{lag}). In
particular, we have demonstrated that even for the ideal dust matter source
there exists a solution {\it without} closed timelike curves and thus the
causality is not violated in such a spacetime.
 
For a specific choice of the teleparallel gravity  parameters,
this solution  reproduces the causality violating stationary G\"odel
solution in general relativity, in accordance with the teleparallel
equivalent of general relativity.
 
Previously it was conjectured that the axial torsion is directly related
to the spacetime rotation, showing that such an assumption is valid for
the case of the Kerr geometry \cite{kerr}. It is interesting to check
whether this is also true for the rotating cosmology under consideration.
 
Using the decomposition of the torsion into three irreducible parts under
the group of global Lorentz transformations \cite{xa}, we can straightforwardly
calculate the axial-vector part
\begin{equation}
A^\mu = {1\over 6}\eta^{\mu\nu\rho\sigma}T_{\nu\rho\sigma}\,.\label{pt3}
\end{equation}
Here $\eta_{\mu\nu\rho\sigma}$ is the completely antisymmetric Levi-Civita 
tensor. Explicitly, we find from (\ref{ga}):
\begin{eqnarray}
A^t = 0,\qquad A^x &=& -\,{\frac 2 {3a}}\left(s_1 - v_3\sqrt{\frac \sigma
{k + \sigma}}\right),\qquad A^y = -\,{\frac 2 {3a}}\,{\frac {e^{-mx}}
{\sqrt{k + \sigma}}}\,s_2,\\
A^z &=& -\,{\frac 2 {3a}}\left(s_3 + v_1\sqrt{\frac \sigma {k + \sigma}}\right)
- \frac{m}{3a^2}\sqrt{\frac{\sigma}{\sigma + k}}.
\end{eqnarray}
For the exact solution obtained, using the equations (\ref{s1}) and (\ref{s3}),
we prove that the first three components vanish $A^t = A^x = A^y = 0$, whereas
\begin{equation}
A^z = -\,{\frac {m\,(2c_1 + c_2)\,(\sigma + 2k)}
{3a^2(2c_1 - c_2)\sqrt{\sigma}\sqrt{k + \sigma}}}\,.\label{Az}
\end{equation}
As we can see, the axial vector torsion is, indeed, related to the spacetime
vorticity. In particular, the axial torsion is apparently directed along the
axis of the cosmic rotation and its value is proportional to the magnitude
of the vorticity. However, this relation should not be overestimated. Let us
recall that $k = -\,\sigma/2$ in the original G\"odel model. Then we find
{}from (\ref{Az}) that the axial torsion vanishes completely, $A^\mu = 0$,
for this case. The existence of the rotating cosmological models with the
vanishing axial torsion shows that one cannot reduce the description of the
cosmic vorticity to the properties of the irreducible parts of the
Weitzenb\"ock torsion tensor.

The earlier studies of the teleparallel gravity have revealed the 
one-parameter family of models which are compatible with the observational 
data \cite{xa,mhn,nester,hecht}. Another characteristic feature of that
family is the existence of black hole solutions which seem to be absent in
the generic case \cite{opp}. On the other hand, exactly this family appears 
to be plagued with the problem of ghost and negative-energy modes as well 
as of the ``hidden" gauge transformations \cite{mhn,kopc,nester,hecht}, see 
also a recent study in \cite{blag}. In this sense, it is interesting that 
the nontrivial spatial axial-vector torsion (\ref{pt3}) of our solution 
prevents the occurrence of the special ``hidden" gauge transformation 
\cite{nester} of the torsion. At the same time, it is worthwhile to note
that the one-parameter family admits only the causality violating cosmological
solutions with rotation, whereas the timelike closed curves are absent in
the generic case. 
 
{\bf Acknowledgments}.
For YNO this work was supported by the Deutsche Forschungsgemeinschaft
(Bonn) with the grant HE~528/20-1. TV would like to thank FAPESP grant 
02/04199-1 for financial support.

\end{document}